# Molecular engineering of chiral colloidal liquid crystals using DNA origami

Mahsa Siavashpouri[1], Christian H. Wachauf[2], Mark J. Zakhary[1], Florian Praetorius[2], Hendrik Dietz[2,3*], Zvonimir Dogic[1*]

[1]Department of Physics, Brandeis University, Waltham, Massachusetts 02454, USA.

[2]Physik Department and Walter Schottky Institute, Technische Universität München, Am Coulombwall 4a, 85748 Garching, Germany.

[3]Institute for Advanced Study, Technische Universität München, Lichtenbergstraße 2a, 85748 Garching, Germany.

*co-corresponding authors

**Establishing precise control over the shape and the interactions of the microscopic building blocks is essential for design of macroscopic soft materials with novel structural, optical and mechanical properties. Here, we demonstrate robust assembly of DNA origami filaments into cholesteric liquid crystals, 1D supramolecular twisted ribbons and 2D colloidal membranes. The exquisite control afforded by the DNA origami technology establishes a quantitative relationship between the microscopic filament structure and the macroscopic cholesteric pitch. Furthermore, it also enables robust assembly of 1D twisted ribbons, which behave as effective supramolecular polymers whose structure and elastic properties can be precisely tuned by controlling the geometry of the elemental building blocks. Our results demonstrate the potential synergy between DNA origami technology and colloidal science, in which the former allows for rapid and robust synthesis of complex particles, and the latter can be used to assemble such particles into bulk materials.**

**Introduction:** The quantitative relationship between the macroscopic properties of a material and the microscopic structure of its constituent building blocks plays an essential role in material science. The promise of colloidal science lies in its ability to continuously tune the shape and interaction of the microscopic building blocks, thus allowing for a rational engineering of hitherto unknown macroscale materials. In this vein, colloids have originally been used to explore the behavior of spheres, rods and platelets that interact through universal hard-core repulsive interactions[1,2]. More recently the focus has shifted towards studying particles with more complex interactions and shapes, and to designing patchy particles that have specific directional attractions[3-9]. However, significant hurdles remain in our effort to elucidate how emergent properties of macroscopic assemblages are determined by the structure and interactions of the constituent colloidal units. A significant obstacle remains a lack of rational method that allows for synthesis of colloidal particles of arbitrary geometry and interactions. Creating any single colloidal architecture typically requires significant effort, and frequently these are not easily generalized.

In comparison, work over the past few years have elucidated principles that allow for rapid and robust assembly of three-dimensional DNA origami particles of almost arbitrary structural complexity[10-12]. Emerging technologies have also enabled the design of highly-tunable attractions based on shape complementarity[13]. While these advances have enabled rational design of individual DNA origami particles, organizing such structures into macroscopic materials remains a significant challenge. A potential synergy between colloidal science and DNA origami technology has thus far remained mostly unrealized[14]. In particular, DNA origami could allow for rapid design of colloidal particles of almost arbitrary geometry and interactions, while methods of colloidal science could provide a robust pathway for assembling origami particles into macroscopic materials.

Here, we explore the unique opportunities that arise at the intersection of colloidal science and DNA origami. We design and purify origami filaments with predetermined aspect ratio and microscopic twist, demonstrate that such filaments robustly form long-lasting cholesteric liquid crystals, and establish a quantitative link between the emergent cholesteric pitch and the microscopic chirality of the constituent rods. Furthermore, we use established methods from colloidal science to assemble DNA origami filaments into 2D colloidal membranes as well as 1D twisted ribbons[15-17]. The latter assemblages behave as effective 1D supramolecular polymers whose mechanical, structural and chiral properties can be engineered by changing the properties of the constituent filaments. Our work outlines a broadly applicable strategy for constructing macroscopic materials that integrates DNA origami technology with self-assembly methods developed for conventional colloids.

**Bulk liquid crystalline behavior of origami filaments:** We have prepared monodisperse DNA origami based particles that mimic the geometry of filamentous phages. Specifically, we synthesized a macroscopic quantity of straight 6-helix (s-6h) origami filaments with 6 nm diameter (D), 420 nm contour length (L), 2.4 μm persistence length ($l_p$) and polydispersity below 3% (see Methods)[18]. Below 28 mg/ml the suspension of s-6h filaments were entirely isotropic. At higher concentrations the sample formed birefringent spindle-shaped droplets suspended in an isotropic background[19]. Over time these sedimented, leading to bulk phase separation between an isotropic and a liquid crystalline phase (Fig. 1c). With increasing concentration, the fraction of the liquid crystalline phase increased, and above 37 mg/ml the samples were uniformly birefringent. These results are consistent with the previous observation of a birefringent fluid of end-fused origami filaments[20]. To elucidate the exact nature of the liquid crystalline order, the origami suspensions were confined within a cylindrical capillary. Origami particles exhibit planar anchoring to capillary walls, which allowed samples to form a fingerprint texture, a hallmark of chiral nematics (Fig. 1a, Methods)[21]. The measured cholesteric pitch continuously decreased with increasing filament concentration, as observed in other lyotropic systems (Fig. 1d)[21,22].

We compared the phase behavior of the 6-helix origami filaments to monodisperse rod-like viruses[23,24], and in particular to the litmus phage which has similar geometry (L=385 nm, D=6.6 nm, $l_p$=2.8 μm)[25]. To account for different particle densities, we have compared the volume fractions of the coexisting isotropic and cholesteric phases ($c_{iso}$, $c_{nem}$). For s-6h filaments at 260 mM ionic strength we estimate, $c_{iso}$=0.04, and $c_{nem}$=0.057 while for Litmus rods at 110 mM ionic strength $c_{iso}$=0.038 and $c_{nem}$=0.05 (see Methods). The influence of the electrostatic repulsion on the isotropic-nematic phase can be taken into account by rescaling the effective filament diameter $D_{eff}$[2]. Decreasing ionic strength from 260 to 110 mM increases the effective diameter from 9.2 to 10.5 nm[2,26]. In the Onsager theory the effective volume fraction for the onset of the nematic phase scales $c_{eff}$=$D_{eff}$/L. Since $L_{DNA}$=1.09$L_{Litmus}$ and $D_{eff(DNA)}$=0.88$D_{eff(Litmus)}$ theory predicts that $c_{eff(Litmus)}$/$c_{eff(DNA)}$=1.23, which is close to what is observed experimentally ($c_{eff(Litmus)}$/$c_{eff(DNA)}$~1.0). The phase behavior of filamentous viruses is quantitatively described by the Onsager theory once it is extended to account for the electrostatic interactions and the filament flexibility[26]. Such quantitative comparison demonstrates that DNA origami filaments interact primarily through repulsive interactions, and establishes them as a model system of rod-like colloids with a number of unique features.

**Tuning chirality of origami based liquid crystals:** DNA origami technology offers the possibility of systematically changing the geometry of the DNA filaments with nanometer resolution. Using this capability we investigated a long-standing, yet still unresolved question in liquid crystals, namely the relationship between the microscopic chiral structure of the constituent filaments and the macroscopic twist of a bulk cholesteric phase[22,27-32]. To this end, we prepared four variants of six-helix bundles with varying degrees of twist along the filament's long axis[10]. In addition to s-6h, we have purified right-handed six-helix (rh-6h) filaments, as well as left-handed (lh-6h) and 2x-left-handed (2x-lh-6h) filaments. Based on a finite-element-based structure prediction, we expect these structures to have, respectively, negligible twist, 360° of right-handed, and 360° and 720° of the left-handed twist inscribed into the structure (Fig. 1b)[33]. Within experimental error, the coexisting isotropic and cholesteric concentrations of all the filaments were the same, in agreement with previous work which demonstrated that Onsager theory describes equally well the isotropic-to-nematic and isotropic-to-cholesteric transition[26].

However, we found that the twist inscribed along the filament's long-axis significantly influenced the pitch of the bulk cholesteric phases (Fig. 1a). Straight 6-helix filaments formed a right-handed cholesteric phase. For same ionic strength and particle concentrations, filaments with 360$^o$ inscribed right-handed twist (rh-6h) formed a more tightly twisted (smaller pitch) cholesteric phase. In comparison, filaments with 360$^o$ of left-handed twist formed cholesterics with larger pitch. Rather surprisingly, further increasing the left-handed twist (2x-lh-6h) yielded a very tightly twisted cholesteric phase (inset Fig. 1d). More detailed analysis described below demonstrates that these filaments formed a cholesteric phase with the opposite handedness, suggesting that filaments with internal twist somewhere between 360° and 720° will form an achiral nematic phase.

Mixing two filaments with opposite chirality leads to the formation of a liquid crystalline phase with tunable macroscopic chirality. We investigated the behavior of binary mixtures by mixing origami filaments with the opposite handedness. Lh-6h and rh-6h filaments, which form cholesteric phase with one chirality, were mixed with 2x-lh-6h filaments, which has the opposite handedness. The pitch of the cholesteric phase was measured as a function of the ratio of the two components (Fig. 1e-f). In contrast to conventional mixtures, where the measured pitch always exhibits linear dependence on the ratio of two species, we found that cholesteric pitch of mixed origami samples exhibited surprisingly complex dependence on the sample composition (Fig. 1e-f). A number of theoretical models predict the relationship between cholesteric pitch and microscopic chirality of the rodlike constituents[30-34]. Our experiments link microscopic and macroscopic chirality. Thus they can be used to test existing theoretical models and guide their future development.

Previous work has demonstrated that the magnitude of the cholesteric pitch and even its handedness is influenced by the slightest structural changes of the constituent filaments, such as switching a single base pair in DNA based liquid crystals, or a single amino acid in filamentous phages[32,34]. However, determining how changes in the chemical composition of the elementary moieties affect the macromolecular structure remains a significant challenge. In comparison, the structure of the DNA origami filaments can be continuously tuned. Our results on bulk-liquid crystals suggest that origami filaments interact through well-understood colloidal interactions, thus paving the way towards systematic exploration of how a particle's shape influences its assembly pathways.

**Structure and mechanics of 1D twisted ribbons:** The experiments described above demonstrate that the interactions between DNA origami filaments are similar to those between filamentous viruses, thus opening up the possibility of using methods of colloidal self-assembly to create new DNA-origami-based materials[15]. In this vein, we have mixed dilute gel-purified DNA origami filaments at concentrations far below the isotropic-cholesteric transition (40 µg/ml) with the non-adsorbing polymer dextran (MW 670000, 35 mg/mL), which induces effective interfilament attractions by the depletion mechanism[35]. Within minutes of sample preparation, such interactions promoted lateral association of filaments and the formation of mesoscopic seed-like structures consisting of a single monolayer of aligned filaments. Subsequently, the seeds coalesced laterally to form a percolating network of 1D twisted ribbons up to hundred microns long that spanned the entire reaction vessel (Fig. 2a). Similar structures were also observed in a mixture of non-adsorbing polymer and filamentous viruses[16,24].

To elucidate the microscopic structures of the 1D ribbon-like assemblages, we employed Differential Interference Contrast (DIC) microscopy, fluorescence microscopy and quantitative LC-PolScope (Fig. 2b-d)[34]. In particular, the LC-PolScope provides 2D spatial maps in which the intensity of each pixel is related to the local optical retardance magnitude and orientation. The LC-PolScope images of 1D ribbons reveal periodic intensity modulation along its contour length. The dark regions correspond to the sections of the ribbon where rods point perpendicular to the image plane. There is little structural anisotropy along this direction, hence very low birefringence. In comparison, bright regions correspond to the sections of the ribbon where rods lie in the image plane and thus exhibit strong structural and optical anisotropy. In this manner LC-PolScope images clearly demonstrate the periodic twist along the ribbon's long axis. DIC imaging reveals a modulation with the same periodicity in the apparent ribbon width, while this structural feature is not easily recognized in fluorescence images. From optical micrographs we were able to reconstruct the 3D structure of twisted ribbons (Fig. 2e). Z-stacks of 2D LC-PolScope images reveal that s-6h, lh-6h and rh-6h filaments formed twisted ribbons with the same handedness whereas ribbons comprised of 2x-lh-6h filaments have the opposite handedness, in agreement with experiments on bulk cholesteric liquid crystals (Fig. 2f-i). These experiments demonstrate that controlling the shape of the microscopic building blocks via DNA origami enables control over the mesoscopic chiral structure of twisted ribbons.

Twisted-ribbons provide a pathway for assembly of 1D supramolecular polymers from a suspension of chemically homogeneous rods that is fundamentally different from the conventional assembly of worm-like micelles comprised of chemically heterogeneous amphiphilic molecules[36]. In order to explore the robustness of this assembly pathway and determine how the properties of supramolecular polymer-like twisted ribbons are influenced by the geometry of the constituent building blocks, we have prepared twisted ribbons from a diverse set of filamentous particles. Besides the origami filaments that were described in the previous section, we have specifically designed and purified additional DNA origami filaments of varying aspect ratio including a 10-helix bundle (L=250 nm, D=8 nm), a 8-helix bundle (L=310 nm, D=7 nm) and 3 rod-like viruses of varying lengths including Litmus (L=385 nm, D=6.6 nm), *fd-wt* (L=880 nm, D=6.6 nm) and M13KO7 (L=1200 nm, D=6.6 nm) (Fig. 3a-i). All rod-like viruses and longer DNA origami filaments robustly assembled into uniform twisted ribbons whose length could reach hundreds of microns (Fig. 3l-q). In comparison, twisted ribbons assembled from shorter DNA origami filaments competed with amorphous aggregation (Fig. 3j-k). S-6h filaments and its chiral variants (rh-6h, lh-6h) formed twisted ribbons with the same structure (Fig. 3r), indicating that the twisted ribbon pitch and width is independent of the microscopic chirality of the constituents, and is primarily determined by their length. A structural analysis revealed that the pitch of twisted-ribbons scales linearly with the length of the constituent particles (Fig. 3r). The same scaling holds for both DNA origami filaments and rod-like viruses, suggesting that the structure of twisted filaments structures is determined by universal principles. Similar relationship was also found in assemblages of chiral amphiphiles[37].

To relate the structure of twisted ribbon to their mechanical properties, we measured their effective bending rigidity, by imaging conformations of a fluctuating ribbon and measuring their fluctuation spectrum, $a_q$ (Fig. 4a-d, Supplementary Movie 1,2,3,4, see Methods)[38]. For large wavelengths $\langle a_q^2 \rangle$ scales as $1/q^2$ (Fig. 4e), demonstrating that twisted ribbons have an effective bending rigidity. At smaller wavelengths the deviations from the expected scaling indicate that the ribbons are no longer structurally homogeneous. The effective persistence lengths ($l_p = \kappa/k_bT$) of twisted ribbons formed from s-6h, as well as litmus, *fd-wt* and M13KO7 phages were respectively, *20 ± 3.4 μm*, *24 ± 2.7 μm, 100 ± 6.6 μm* and *153 ± 10 μm*, demonstrating that $l_p$ can be tuned by controlling the length of the constituent rods (Fig. 4e).

Besides bending rigidity, the other parameter that characterizes the behavior of 1D twisted ribbons is their extensibility, which is related to the effective Young's modulus of ribbons. To measure the elasticity of twisted ribbons assembled from six-helix DNA origami filaments we have performed force-extension experiments with single ribbons[39]. Briefly, one end of a ribbon was held and stretched with an optical trap while simultaneously the force exerted on the other end was measured (Fig. 4f, Supplementary Movie 5,6). Directly trapping a twisted ribbon with an optical trap exerts a torque and a force. To create a torque-free boundary condition, the optical traps were removed from direct contact with the ribbons by assembling rigid dumbbell handles that are comprised of two micron-sized beads that are connected with a rigid flagellar filament (Fig. 4g). One end of the dumbbell was attached to the ribbon by depletion forces while the optical trap held the other end. Such experimental geometry yielded reproducible force-extension curves (Fig. 4h).

**Phase-diagram of origami filaments:** So far we have discussed the assembly and the mechanics of 1D twisted ribbons formed from DNA origami building blocks. To explore the formation of other self-limited structures, we systematically altered two parameters that control the self-assembly pathways of colloidal rod-like particles, namely the solution ionic strength and the depletant concentration, which determine the range and the strength of the attractive interactions, respectively. By varying these two parameters, we mapped the equilibrium phase diagram for the assembly of origami filaments (Fig. 5a). Comparison with the phase diagram for filamentous viruses reveals similarities but also some notable differences. At low ionic and depletant concentrations, filaments stay in the isotropic phase, however at high ionic strength, strong depletion attractions drive assembly of disordered aggregates. Most importantly for both *fd* viruses and DNA origami filaments we observed the same three morphologies, namely 1D twisted ribbons (Fig. 5d), 2D isolated monolayer membranes (Fig. 5b) and 3D smectic-like stacks of colloidal membranes (Fig. 5c). While viruses form twisted ribbons, monolayer membranes, and stacked membranes with increasing depletant concentration[15,16,40], the sequence of phases for origami filaments is reversed. The microscopic reason for this intriguing inversion remains unclear and requires more detail studies.

The analogy between *fd* viruses and origami filaments extend to the dynamics of the phase transitions. Decreasing chirality increases the membrane edge tension and leads to a polymorphic transitions of 1D twisted ribbons into 2D colloidal membranes (Fig. 6a, Supplementary Movie 7)[16,41,42]. We observe a similar transition in origami-based twisted ribbons. Illuminating a freely fluctuating ribbon with green excitation light nucleates a flat membrane at one of the ribbon ends. Once nucleated, the colloidal membrane grows continuously as it absorbs materials from the twisted ribbon portion of the assemblage (Fig. 6b, Supplementary Movie 8). Only the regions of the sample that were exposed to the fluorescent excitation light exhibit the photo-induced polymorphic transition, suggesting that the ethidium-bromide (EtBr) DNA intercalating stain which was also present in the sample, plays an important role in controlling the microscopic chirality of origami filaments. The effect of UV light on the intercalation of EtBr and therefore the chirality of DNA origami filaments is not well understood at this point.

**Discussion and Conclusions:** We have demonstrated methods for assembly of diverse origami based structures that are complementary to existing methods of assembling DNA based liquid crystals[43,44]. All origami based materials exhibited long-term stability that lasted more than a year. Our results are important for several reasons. First, they establish origami filaments as a promising model system to study the self-assembly of soft materials with orientational order. Tuning aspect ratio and various geometrical features of origami particles is easily accomplished. The origami technology could be used to create more exotic colloidal particles and study how they assemble into macroscopic materials. Second, our work on DNA origami filaments demonstrates that assembly of colloidal membranes is a ubiquitous feature of rod-polymer mixtures, as has been suggested by theoretical models[40]. Original work demonstrating assembly of twisted ribbons and colloidal membranes was carried out with viruses. Observation of a similar assembly pathway in a different system shows that the stability of colloidal membranes is a consequence of universal excluded volume interactions[40,45]. Third, using both DNA origami filaments and rod-like viruses we have assembled 1D twisted ribbons, a ubiquitous structural motif observed in biology and chemistry, ranging from amyloid fibrils to cholesterol crystallization in human bile[46-48]. We have demonstrated that the twisted ribbons effectively behave as 1D supra-molecular polymers, whose mechanical and structural properties can be tuned by controlling the geometry and chirality of the constituent filaments. The classical paradigm for assembly of 1D supra-molecular polymers is based on assembly of worm-like micelles from amphiphilic building blocks[49], in which the filament stiffness is tuned by controlling the size of the two incompatible segments. Our work demonstrates an alternate route towards assembly of designable 1D polymer-like structures from chemically homogeneous monodisperse rods. Finally, our methods could be extended to assemble origami-based chiral plasmonic nanostructures into hierarchical materials that might exhibit intriguing photonic properties[50].

**Acknowledgments:** We acknowledge support of NSF-MRSEC-1420382 and NSF-DMR-1609742 (to M. S., M. J. Z. and Z. D). We also acknowledge use of Brandeis MRSEC optical microscopy and biosynthesis facility supported by NSF-MRSEC-1420382. This work was also supported by a European Research Council Starting Grant to H.D. (GA no. 256270) and by the Deutsche Forschungsgemeinschaft through grants provided via the TUM Institute of Advanced Study, the Cluster of Integrated Protein Science, the Nano Initiative Munich, and the Gottfried-Wilhelm-Leibniz Program (C.H.F., P.F. and H.D.).

**Author contributions:** H.D. and Z.D. conceived the experiments. M.S. and M.J.Z. performed initial experimental observations. M.S. performed all the experiments. C.H.W. designed and characterized origami filaments. Z.D, M.S, H.D, C.H.W. and M.J.Z. analyzed the experimental data. F.P. developed methods to purify large quantities of scaffold and provided them for our studies. M.S, H.D. and Z. D. wrote the manuscript. All authors revised the manuscript.

**Competing financial interests:** The authors declare no competing financial interests.

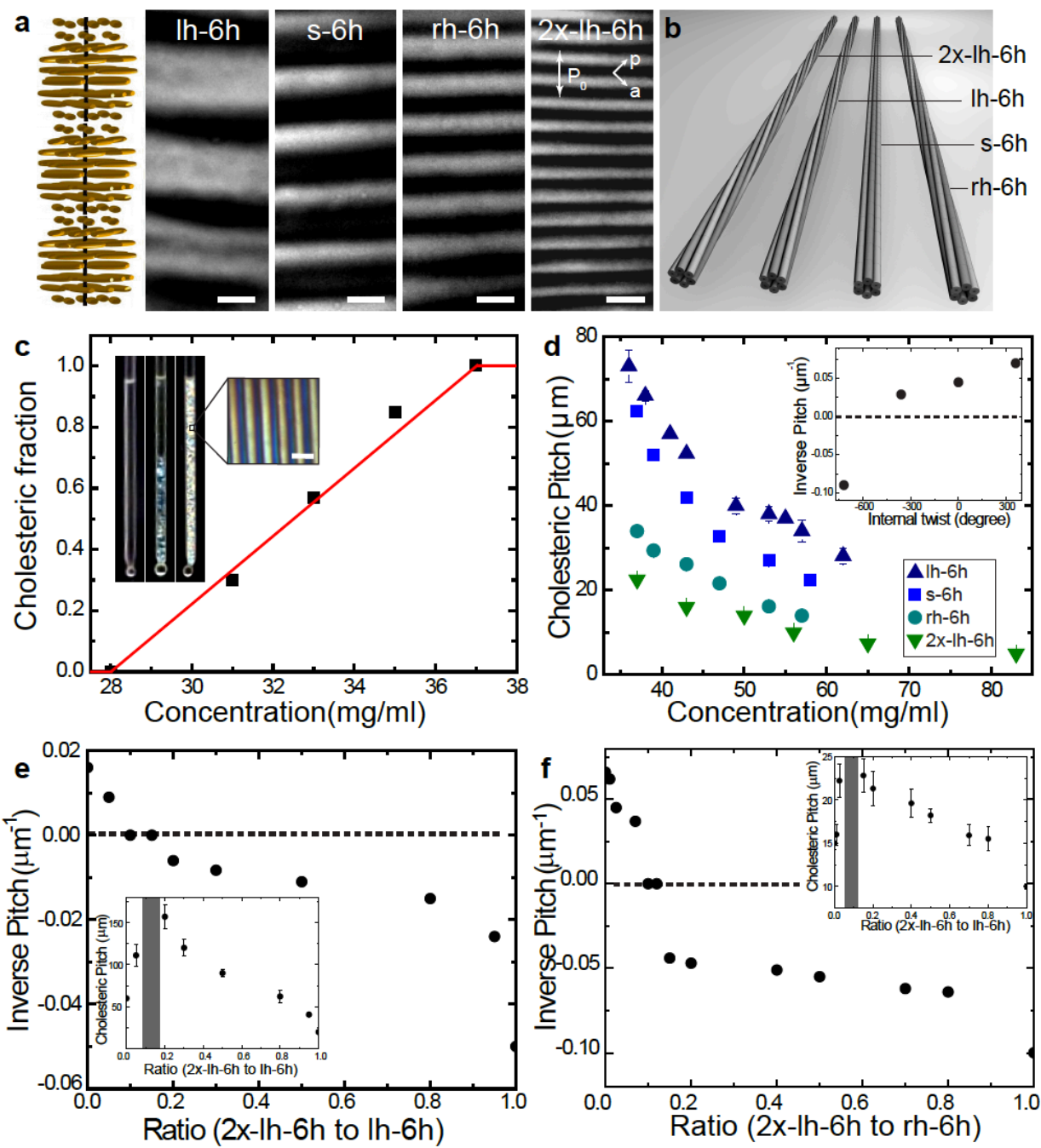

**Figure 1. Origami based cholesteric liquid crystals: a,** Schematics and polarization micrograph of cholesteric liquid crystalline samples comprised of lh-6h, s-6h, rh-6h and 2x-lh-6h filaments. The concentration of all cholesteric samples is 37 mg/ml and ionic strength is 260 mM. Dark lines represent the regions where the rods are perpendicular to the image plane and bright lines correspond to regions where rods are in the image plane. The length of the cholesteric pitch ($P_0$) and orientation of the polarizer (p) and analyzer (a) are indicated in the micrograph. Scale bars, 20 μm. **b,** Structure of four different DNA origami filaments with varying twists along the filament's long axis that are used for assembly of bulk liquid crystals. **c,** The volume fraction of the s-6h cholesteric phase increases with increasing origami filament concentration, demonstrating a first order isotropic-cholesteric phase transition. The plot represents the cholesteric fraction of the sample vs. sample concentration. Inset: The coexistence between isotropic and cholesteric phase of three individual samples in the capillary tubes. From left to right: isotropic phase (C~26 mg/ml), coexisting isotropic-cholesteric phase (C~33mg/ml) and single cholesteric phase (C~40mg/ml) with the fingerprint texture of cholesteric phase, Scale bar, 40 μm. **d,** Cholesteric pitch as a function of origami filaments concentration for four different twist variants. Inset: Plot of inverted pitch ($1/P_0$), which is proportional to the chiral twist constant, as a function of internal twist of DNA origami filaments. Increasing left-handed twist increases cholesteric pitch eventually switching the handedness of the cholesteric phase. **e,** Inverse pitch of bulk liquid crystals is plotted as the ratio of 2x-lh-6h to lh-6h filaments. The overall concentration is 40 mg/ml. Pitch was larger than sample size at ratios between 5% to 15% of 2x-lh-6h. Inset: Cholesteric pitch as a function of the filament ratio. **f,** Plot of inverse pitch as a function of 2x-lh-6h to rh-6h filaments ratio. The overall concentration is fixed at 55 mg/ml. Cholesteric pitch could not be measured between 7% to 12% of 2x-lh-6h. Inset: Cholesteric pitch as a function of filament ratio.

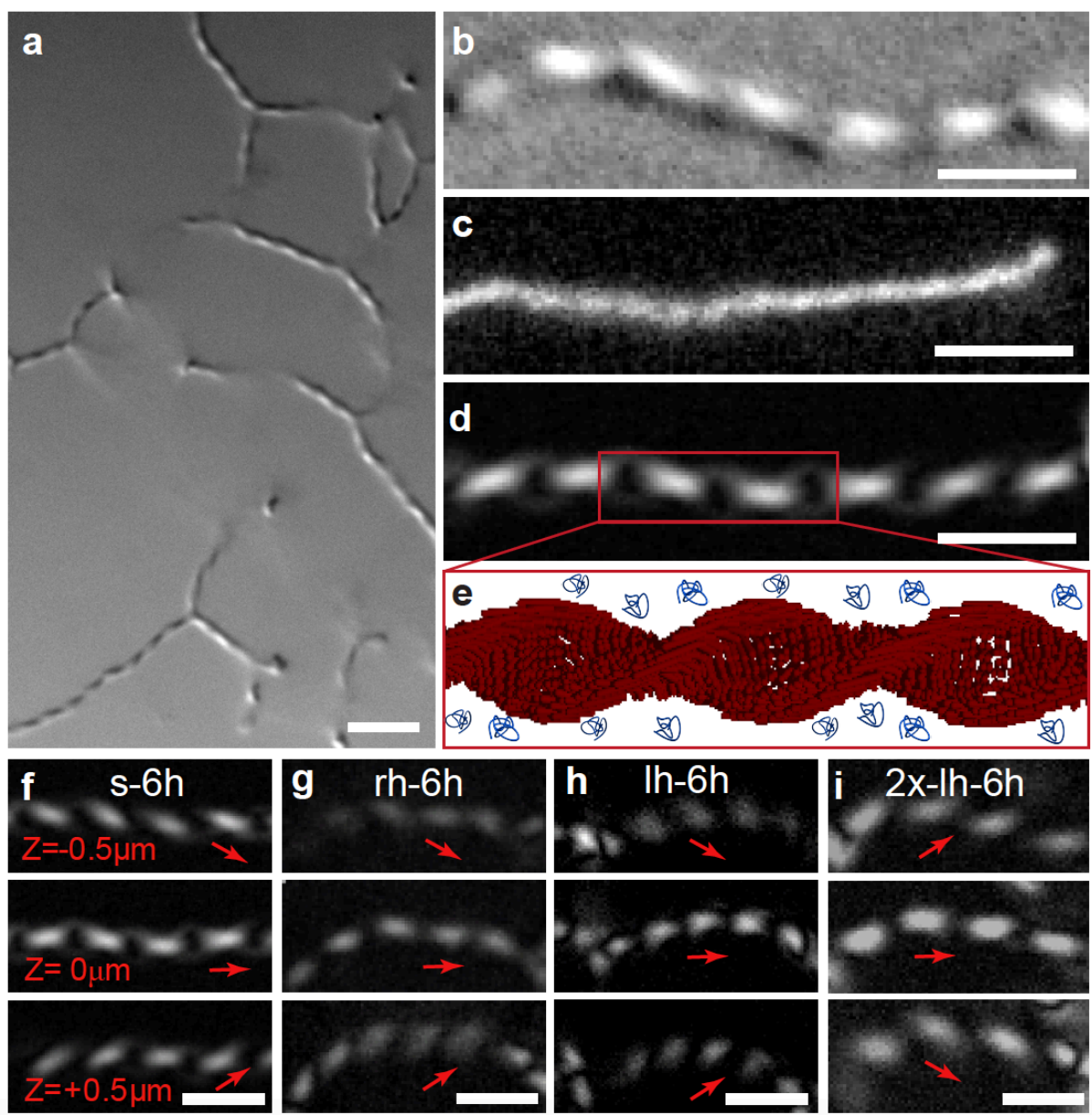

**Figure 2. Origami filaments assemble into supramolecular 1D twisted ribbons. a,** Non-adsorbing polymer induces DNA origami filaments to assemble into a percolating network of 1D twisted ribbons. Scale bar, 4 μm. **b-d,** Differential Interference Contrast (DIC; b), fluorescence (c) and LC-PolScope (d) images demonstrating the periodic nature of a twisted ribbon. In the LC-PolScope image the dark regions correspond to the ribbon sections where rods point

perpendicular to the image plane, and bright regions correspond to the sections of the ribbon were rods lie in the image plane. Scale bars, 2 μm. **e,** Schematics of 1D twisted ribbon in suspension of non-absorbing polymers. **f-i,** Z-stacks of the LC-PolScope images determines the handedness of twisted ribbons (as indicated by the red arrows). s-6h (f), rh-6b (g) and lh-6h (h) filaments assemble into the left-handed ribbons while 2x-lh-6h filaments (i) form right-handed ribbons. Scale bars, 2 μm.

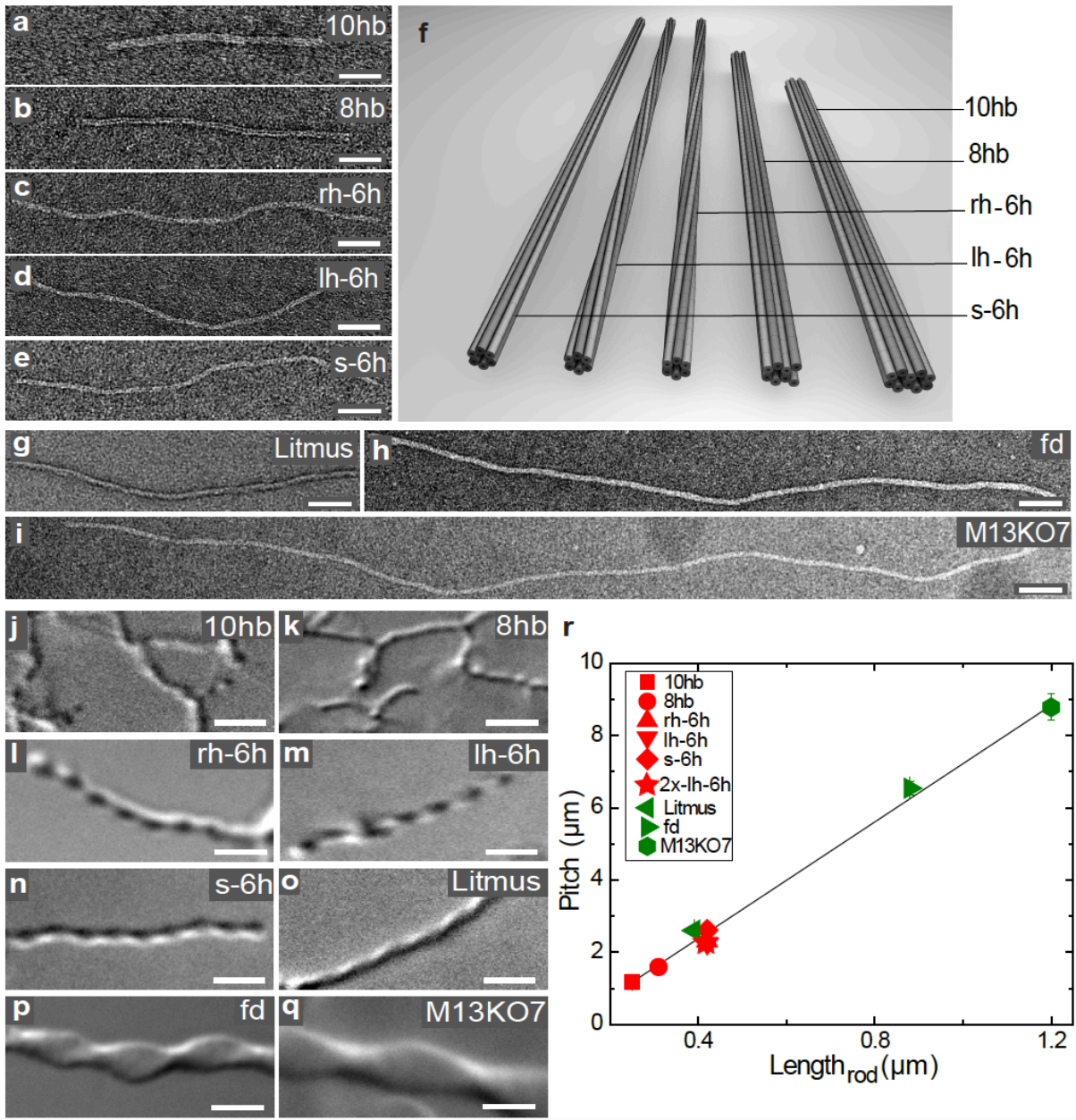

**Figure 3. Geometry of the constituent rods alone determines twisted ribbon structure. a-f,** Electron-microscope micrographs and schematic models of a 10 helix filament (a; L=250 nm, D=8 nm), a 8 helix filament (b; L=310 nm, D=7 nm), a right-handed 6-helix filament (c; L=420 nm, D=6 nm), a left-handed 6-helix filament (d; L=420 nm, D=6 nm) and a straight 6-helix filament (e; L=420 nm, D=6 nm). Scale bars, 50 nm. **g-i,** TEM micrographs of Litmus (g; L=385

nm, D=6.6 nm), *fd-wt* (h; L=880 nm, D=6.6 nm) and M13KO7 (i; L=1.2 μm, D=6.6 nm) filamentous bacteriophages. Scale bars, 50 nm. **j-q,** DIC micrographs of 1D twisted ribbons self-assembled from origami filaments and rod-like viruses shown in a-i. Scale bars, 2 μm. **r,** The pitch of twisted ribbons increases linearly with the length of the constituent filaments. The same scaling observed for origami and virus twisted ribbons, suggests that the assemblage structure does not depend on the chemistry of the elemental rods but only on their geometry.

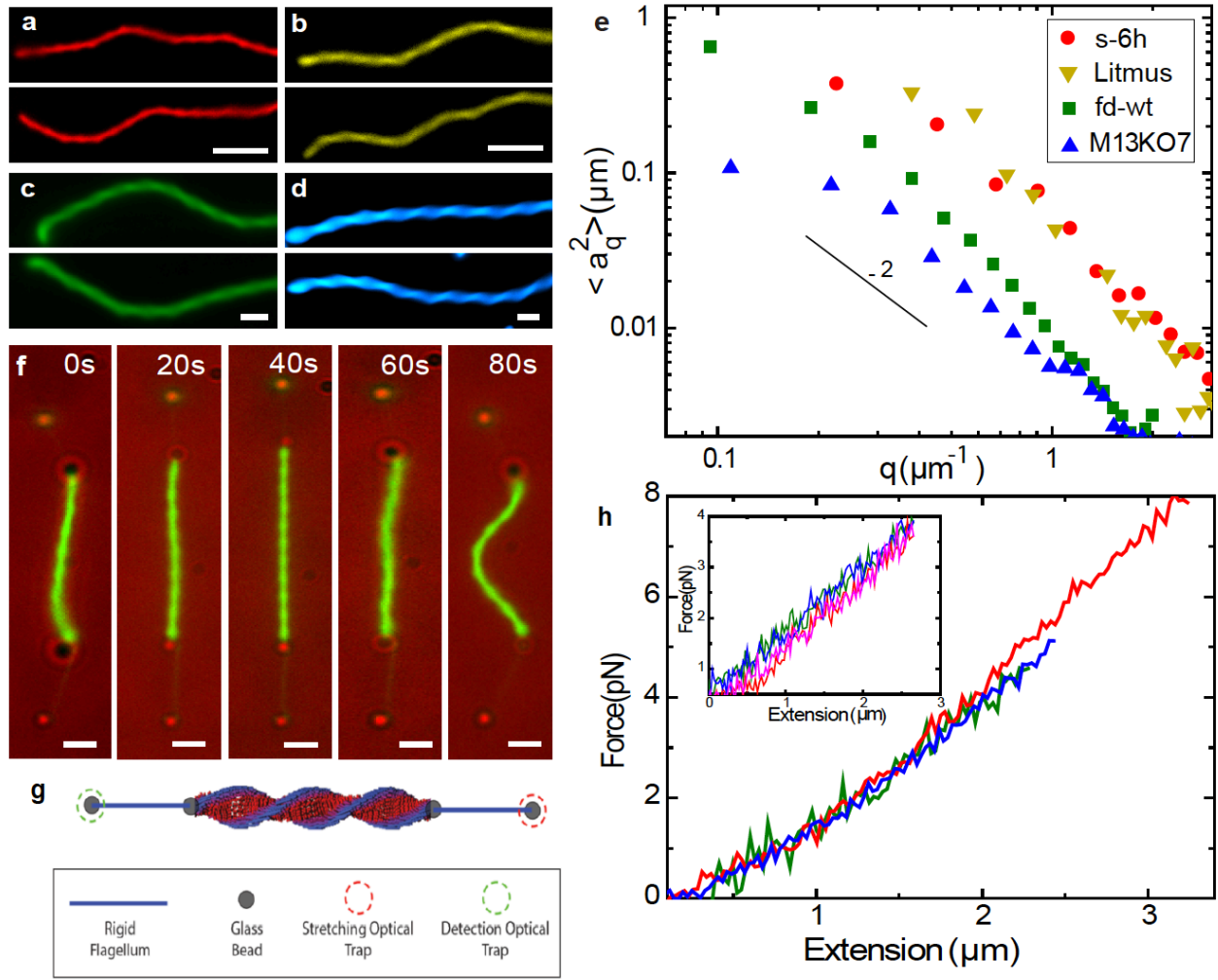

**Figure 4. Mechanical properties of 1D twisted ribbons. a,** Fluorescence micrographs of two different configurations of a twisted ribbon self-assembled from s-6h filaments which exhibits significant thermal fluctuations, implying a relatively low effective bending rigidity. Scale bar, 4 μm. **b-d,** Fluorescence images of two random configurations of Identical ribbons assembled from Litmus virus particles (b), *fd-wt* filaments (c), and M13KO7 bacteriophage (d). Scale bars, 4 μm. **e,** Fluctuation spectrum of s-6h origami, Litmus, *fd-wt* and M13KO7 twisted ribbon. At low wavenumbers (*q*), fluctuation spectrums scale as $q^{-2}$ as is expected for semi-flexible filaments allowing us to extract the effective bending rigidity. **f,** A time sequence showing a force-extension of a twisted ribbon assembled from fluorescently labeled s-6h filaments. The twisted

ribbon is stretched by using rigid dumbbells as two handles. Brightfield channel (red) shows four glass beads forming dumbbell-like structures, while the fluorescence channel (green) shows the fluorescently labeled s-6h DNA origami twisted ribbon. Scale bars, 2 μm. **g,** A schematic illustration of the experimental setup for the force-extension experiments of a twisted ribbon assembled from s-6h filaments. Dumbbell-shaped objects are created from glass beads bound to rigid, linear flagella. Two dumbbells are attached by depletion interactions to two ribbon ends, respectively. The ribbon is stretched from one end using an optical trap, while the response force exerted by the stretched ribbon is extracted from the displacement of the bead from the detection trap at the other end. **h,** Force-extension measurements on multiple twisted ribbons indicate the effective extensibility of twisted ribbons. Inset: Repeating extension and relaxation cycles of a single ribbon demonstrate the reproducibility of the force-extension measurements.

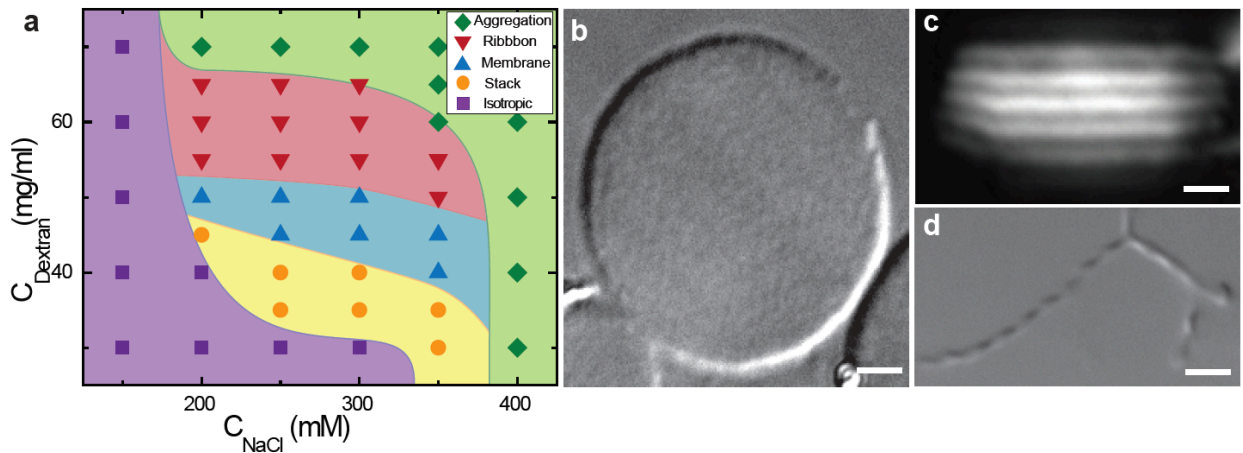

**Figure 5. Phase diagram of an origami-depletant mixture. a,** The experimental phase diagram of s-6h DNA origami filaments as a function of depletant concentration and ionic strength. The concentration of DNA origami filaments was fixed at 40 μg/ml. At low ionic strength the repulsive interactions dominate over depletion attractions for all dextran concentrations and DNA origami filaments remain suspended in an isotropic phase (purple). At high ionic strength, strong depletion attractions drive assembly of disordered aggregates (green). Intermediate ionic strength favors assembly of well-defined structures such as membranes and ribbons. Specifically, twisted ribbons are energetically favored at higher dextran concentration (red). Reducing the osmotic pressure (depletant concentration) induces a transition into 2D colloidal membranes comprised of a one rod-length thick monolayer of aligned nanorods (blue). Decreasing the dextran concentration further leads to a second transition where 2D membranes stack on top of each other to form bulk 3D smectics (yellow). **b,** DIC micrograph of a 2D isolated colloidal monolayer membrane (face-on) observed at intermediate depletant concentrations and ionic strength. **c,** Fluorescence image of 3D smectic-like stacks of colloidal membranes observed at low depletant concentrations. **d,** A DIC micrograph of 1D twisted ribbon formed at high dextran concentrations. Scale bars, 2 μm.

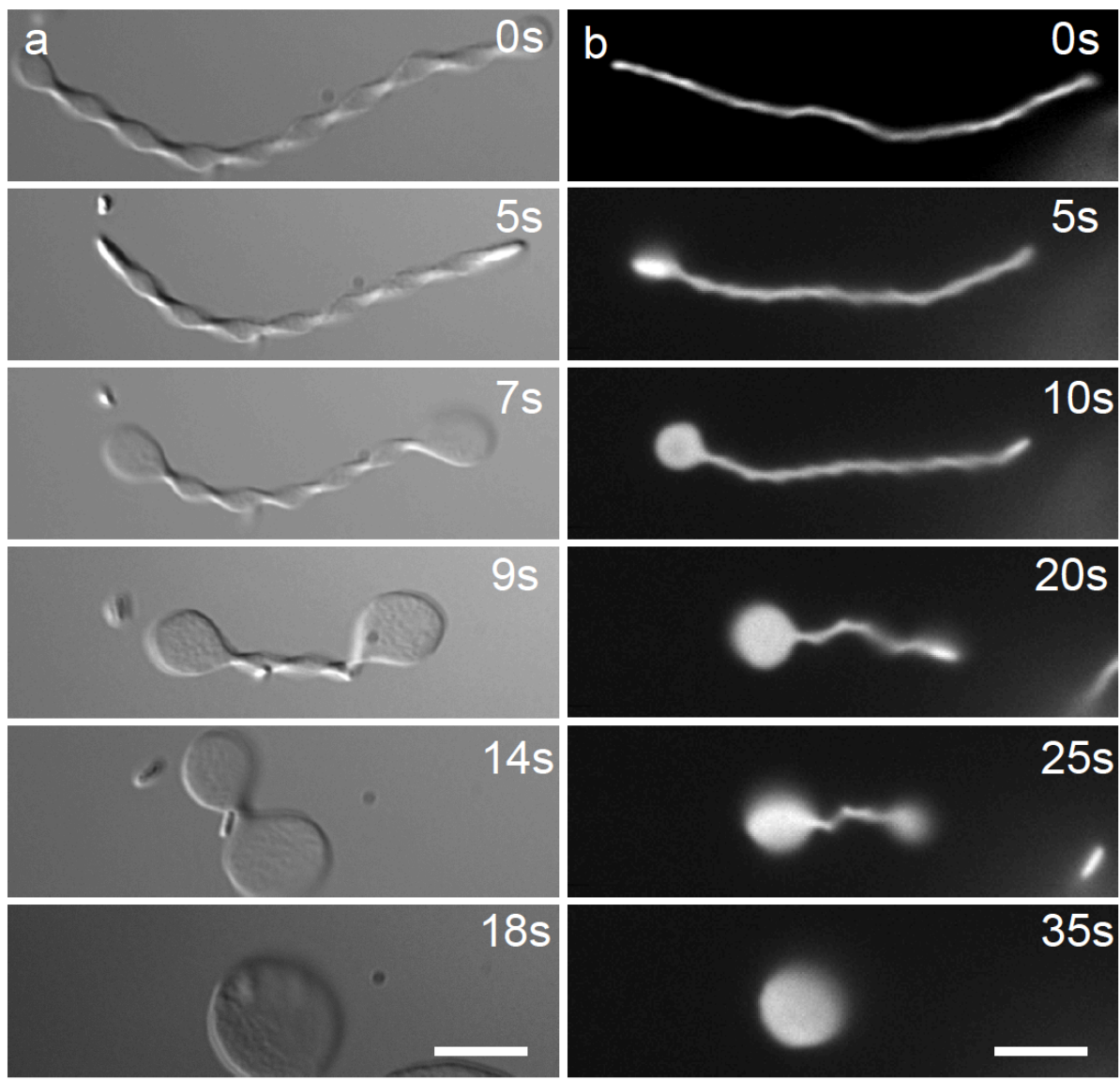

**Figure 6. Stimuli induced ribbon to membrane transition. a,** A sequence of DIC images illustrates polymorphic transition of 1D ribbons into flat 2D colloidal membranes. The ribbon is comprised of *fd-wt* and with increasing temperature (decreasing *fd-wt* chirality) undergoes a polymorphic transition. **b,** A similar ribbon-to-membrane transition observed in a suspension of s-6h DNA origami filaments. The transition is induced by exposure to fluorescence excitation light. Scale bars, 4 µm.

## Methods:

**Design of origami filaments:** Origami filaments are comprised of parallel DNA double helices that are connected by crossover sites[51]. To suppress aggregation associated with blunt-ends, polythymidine tails and single-stranded scaffold DNA loops at the helical interfaces were introduced[52]. The single-stranded 7560 bacteriophage genome has been used as a template for DNA origami. Phage 7560 has been grown, harvested and purified as previously described[53]. s-6h DNA origami objects are assembled by designing 177 short ssDNA called "staples" with the average length of 42 base pairs. caDNAno program[54] was used for designing origami objects. Individual staples have range in length from 21 to 49 nucleotides. 177 staple oligonucleotides were mixed to make a working stock where each molecule was at a standard concentration of 500 nM. In addition to the straight 6-helix filaments (s-6h), we designed variants of 6-helix bundles with differing handedness by insertion and deletion of base pairs along the bundle contour as described elsewhere[10]. Furthermore we also designed multi-layer DNA origami filaments including 8 and 10 helices.

**Material yields and cost of preparing origami filaments.** The staple strands cost is about $1200 per 10 mg of origami filaments. For the purpose of every bulk liquid crystal experiment, 4-mg origami filament was prepared at the final concentration of approximately 60 mg/ml. However the final yield of folding is correlated to the filament structure. 2x-lh-6h has the highest yield and rh-6h has the lowest yield of folding. For obtaining a 4-mg yield of origami filaments, about 8 mg ssDNA scaffold was extracted from 45 mg 7560 bacteriophage.

**Large-scale DNA origami folding reaction for the bulk liquid crystal experiments.** 60 ml folding reaction was prepared with a 3:1 excess of each staple over scaffold. Scaffold at an effective concentration of 66.6 nM and staples at 200 nM (staples were extracted from working stock) where mixed in the folding buffer (5 mM NaCl, 5 mM Tris, 1 mM EDTA and 20 mM $Mgcl_2$, pH=8.0). Folding reaction was subjected to an annealing procedure with the following program: 1. 65°C for 15 min, 2. 58°C for 24 min (-1°C per cycle). 3. Repeat 2, 4 times. 4. End

**Large-scale DNA origami purification for the bulk liquid crystal experiments.** Folded origami particles purified from the excess staples and miss-folded objects via the following protocol:

1- Spin at 33000 rcf at 4ºC for 20 minutes

2- Transfer supernatant into fresh container

3- To supernatant, add equal volume of 15% polyethylene glycol ($M_w$8000), 20 mM $MgCl_2$ and 500 mM NaCl mixed in the buffer of 1x FOB (1 mM EDTA, 5 mM Tris base and 5 mM NaCl at pH=8.0).

4- Spin at 16000 rcf at 4ºC for 30 minutes.

5- Discard the supernatant and re-suspend the pellet into buffer of 250mM NaCl, 20mM Tris base at pH=8.0.

6- Spin at 340,000 rcf for 2 hours.

7- Re-suspend the pellet into buffer of 250 mM NaCl, 20 mM Tris base at pH=8.0.

8- Using UV absorbance at 260 nm, the concentration of DNA origami filaments was estimated assuming an extinction coefficient of $A_{260} = 1$ for 50 μg/ml.

**Sample preparation of bulk liquid crystals.** Bulk liquid crystalline phases have been characterized using previously developed methods[21,34]. Boron-rich thin-walled capillary tubes with diameter of 0.7 mm (Charles Supper Company, Natick MA) were filled with high concentration suspensions of origami filaments. Prior to filling the sample, capillary tubes were cleaned in a hot detergent solution (Hellmanex), soaked in 5 M NaOH and subsequently cleaned with deionized water. The cleaning process ensured parallel anchoring of origami liquid crystals to the capillary walls. After filling the sample, tube was sealed properly with miniature blowtorch. Within few hours, the sample equilibrated exhibiting the characteristic fingerprint texture under polarization microscopy. These samples are stable at least few years when stored at 4ºC.

The DNA origami filaments show planar anchoring to a planar glass surface, which has the same chemistry as the cylindrical glass capillary tubes used in our experiments. While the capillary tube has no curvature along its long axis, it has a finite curvature perpendicular to its long axis. The curvature breaks the in-plane degeneracy and causes a preference in alignment of the DNA origami filaments at the capillary interface along the long axis of the cylinder. As a consequence, the cholesteric pitch is perpendicular to the cylinder's long axis. The cholesteric fingerprint texture extends about one third into the bulk interior of the capillary. Beyond, the cholesteric phase melts into a nematic phase at the core of the cylindrical capillary to resolve the packing frustration.

**DNA origami folding reaction for the twisted ribbon assembly experiments.** Folding reaction was prepared in a PCR strip tube at total volume of 100 µl with a 5:1 excess of each staple over scaffold. Scaffold was at effective concentration of 40 nM and staples at 200 nM (staples were extracted from working stock), mixed in the buffer of 5mM NaCl, 5mM Tris base, 1mM EDTA and 20 mM Mgcl$_2$ prepared at pH=8.0[33]. Folding reaction then was subjected to a thermal denaturation and annealing procedure with the following program: 1. 65°C for 15 min, 2. 60.0°C for 1 hour (-1°C per cycle). 3. Goto 2, 17 times. 4. End. This protocol was applied for all the DNA origami variants.

**DNA origami purification for the twisted ribbon assembly experiments.** Quality of the folding reactions was examined by running the folded solution into the 2% agarose gel electrophoresis. The gel tray was kept in ice-water bath. 0.3 µg/ml of ethidium bromide was added to the running buffer before electrophoresis. A voltage of 90V was applied for about 2 hours. Folded DNA origami filaments were separated from excess staples and miss-folded particles during a gel isolation procedure. The desired bands were extracted from agarose gel slabs and cut into the smaller pieces by a razor blade, frozen at -20 for 5 min then centrifuged at 16000 rcf for 5 min at 25°C in a freeze'n'squeeze spin column (Biorad). Origami filaments were precipitated by adding 1 volume of 15% PEG (M$_w$8000), 20 mM MgCl$_2$ and 500 mM NaCl mixed in the folding buffer (5 mM NaCl, 5 mM Tris, 1 mM EDTA and 20 mM Mgcl$_2$, pH=8.0). and spinning at 16000 rcf for 30 min. Final origami filaments were re-suspended into the buffer of 250 mM NaCl, 20 mM Tris base pH=8.0 to the desired concentration. . This protocol was applied for all the DNA origami variants.

**Sample preparation of origami-polymer mixtures.** DNA origami filaments were mixed with the polymer dextran (MW 670000, Sigma-Aldrich). The final concentration of DNA origami varied from 30 to 200 ng/µl, the final concentration of polymer varied from 20 mg/ml to 80 mg/ml. Samples were prepared in a buffer solution containing 20 mM Tris base and 300 mM NaCl at pH=8.0. All samples were prepared in an optical microscopy chamber which consists of one glass slide and one coverslip attached together via a layer of unstretched parafilm. In order to prevent nonspecific binding of DNA to the glass slide and coverslip surfaces, an acrylamide treatment was applied[55]. Self-assembled structures form in the bulk in a few hours and sediment to the coverslip due to the high-density effect. Twisted ribbons and colloidal membranes only formed when non-adsorbing polymer is added to the diluted isotropic suspension of DNA origami filaments. At higher concentrations origami particles formed the bulk liquid crystalline phases described above.

**DNA origami labeling:** Two alternate methods were used to fluorescently label origami filaments and their assemblages. In a first method, a Cy3 dye was end-bound to one of the oligonucleotide staples used in a folding reaction. In a second method we used ehidium-bromide as fluorescent marker of large-scale DNA assemblages. Ethidium bromide was present in all experiments that involved assembly of 1D twisted ribbons and 2D membranes since the origami filaments were purified with gel electrophoresis. To reduce photobleaching effects a standard oxygen scavenging solution is added consisting of glucose oxidase, catalase and glucose[56].

**Filamentous Bacteriophage:** Large quantities of virus were grown and purified using standard biological procedures[57]. Virus was mixed with dextran (MW 670 000, Sigma-Aldrich) in a buffer consisting of 100 mM salt and 20 mM Tris HCl at pH=8.0. Final concentration of virus and dextran was 2 mg/ml and 35mg/ml. The optical density of Litmus, *fd wt* and M13KO7 for 1 mg/ml solution is 3.84 at $\lambda$=269 nm. In order to measure bending rigidity, Litmus, *fd-wt* and M13KO7 bacteriophages were labeled by Dylight550 NHS ester (Thermo Fisher Scientific, 62262) based on the previously published protocol[58].

**Microscopy.** Samples were visualized by an inverted microscope (Nikon TE-2000) equipped with a differential interference contrast (DIC) module, a fluorescence imaging module and a 2D-LC-Polscope module[59]. A high numerical aperture oil objective (100X PlanFluor NA 1.3) and a Mercury Halide lamp (Excite-120) were used. Images were collected with a high sensitivity cooled CCD cameras (Andor-Clara for DIC, Polscope and Andor iXon for fluorescence imaging) operating in a conventional mode. In fluorescence imaging, exposure time was kept at minimum (less than 20 ms) to reduce blurring effects.

**Optical tweezers.** A holographic optical tweezer setup was used with a 1,064 nm laser beam (Compass 1064, Coherent). Multiple point traps were created and translated in the image plane in real time. To perform the force-extension measurements on twisted ribbons, an experimental set-up was developed to measure the force required to stretch a ribbon as a function of extension. Thus, we trap both ends of the twisted ribbon with two optical traps, stretch one end with one of the tarps in small steps and measure the force exerted on the other trap at each step. During this experiment, a quadrant photodiode (QPD) position detection was used which allows for nanometer-precision position detection of a spherical bead. A spherical glass bead exhibits a linear response over a displacement of about 0.5μm, which is sufficient for the range of forces associated with the twisted ribbon stretching. Therefore, to maintain the linear displacement-voltage response of QPD detection and to avoid any boundary condition on the rod orientations at the optical trap, dumbbell-shaped objects were attached to both ends of the ribbon and isolated it from both the detection and optical traps (Fig.4e). Each dumbbell-shaped object is comprised of one bacterial flagella and two-micron size glass beads that are attached to either end of the flagella by biotin-streptavidin linkers. Bacterial flagella were isolated from Salmonella typhimurium strains SJW 1660 (straight morphology) as previously described[60]. The dumbbell that is responsible for stretching the ribbon has one bead attached to the ribbon by depletion interactions, while the optical tweezer traps the other bead. The second dumbbell is placed at the detection end of the ribbon where the detection position is performed on the isolated bead and exhibits a linear displacement-voltage response.

**Volume fraction measurements of coexisting isotropic and cholesteric phases ($c_{iso}$, $c_{nem}$):** We measured the volume fractions of DNA origami and virus filaments at the coexisting phases ($c_i$) by following this equation:

$$c_i = (\pi/4)\rho_i L D^2$$

where $\rho_i = \frac{n_i}{V}$ is the number density of filaments of the coexisting phase, $n_i$ is the number of filaments and V is the total volume of the sample. L is the contour length and D is the diameter of filaments.

**Measuring twisted-ribbon persistence length:** Large-scale ribbon fluctuations can be described by the following Hamiltonian:

$$H = \frac{\kappa}{2} \int \left(\frac{d\theta(s)}{ds}\right)^2 ds = \sum_q \frac{\kappa a_q^2 q^2}{2},$$

where $\kappa$ is the effective bending rigidity, and $\theta(s)$ denotes the local curvature along the filament contour length $s$. $\theta(s)$ is decomposed into Fourier components of amplitude $a_q$ according to the following expression: $\theta(s) = \sqrt{2/L} \sum_q a_q \cos(qs)$. It follows that $\kappa = \frac{k_b T}{q^2 \langle a_q^2 \rangle}$, where $\langle a_q^2 \rangle$ is averaged over independent conformations[38].

**Supplementary Movies:**

**Supplementary Movie 1**

Twisted ribbon assembled from fluorescently labeled s-6h origami filaments exhibits significant thermal fluctuations. Scale bar, 4 μm.

**Supplementary Movie 2**

Thermal fluctuations of a twisted ribbon self-assembled from fluorescently labeled Litmus viruses. Scale bar, 4 μm.

**Supplementary Movie 3**

A fluorescently labeled *fd-wt* bacteriophage-based twisted ribbon undergoes thermal fluctuations. Scale bar, 4 μm.

**Supplementary Movie 4**

Twisted ribbon comprised of fluorescently labeled M13KO7 bacteriophages undergoes thermal fluctuations. Scale bar, 4 μm.

**Supplementary Movie 5**

Force-extension experiment performed on s-6h DNA origami-based twisted ribbon via optical trap. Two rigid dumbbells are used as handles to create torque free boundary conditions. One end of the ribbon is stretched with an optical trap while the force exerted on the other end is measured. Scale bar, 2 μm.

**Supplementary Movie 6**

Force-relaxation experiment performed on s-6h DNA origami-based twisted ribbon via optical trap. Two rigid dumbbells are used as handles to create torque free boundary conditions. One end of stretched twisted ribbon is released from the optical trap while the force exerted on the other end is measured. Scale bar, 2 μm.

**Supplementary Movie 7**

Temperature-induced transition of *fd-wt* twisted ribbon into a 2D colloidal membrane. Scale bar, 2 μm.

**Supplementary Movie 8**

A polymorphic transition of 1D twisted ribbon comprised of s-6h DNA origami filaments into 2D colloidal membrane that nucleates at the end of a freely fluctuating twisted ribbon. The transition is induced by exposing to green excitation light. Ethidium-bromide is the marker for fluorescently labeled DNA origami filaments. Scale bar, 2 μm.